# A Critique of Drexler Dark Matter

C. Sivaram, Kenath Arun and R. Nagaraja

Indian Institute of Astrophysics, Bangalore

**Abstract:** Drexler dark matter is an alternate approach to dark matter that assumes that highly relativistic protons trapped in the halo of the galaxies could account for the missing mass. We look at various energetics involved in such a scenario such as the energy required to produce such particles and the corresponding lifetimes. Also we look at the energy losses from synchrotron and inverse Compton scattering and their signatures. The Coulomb repulsive instability due to the excess charge around the galaxies is also calculated. The above results lead us to conclude that such a model for DM is unfeasible.

Dark matter (DM) is supposed to constitute about 90% of the mass of typical galaxies and about a third of the density of the universe [1, 2]. Several candidates have been proposed for dark matter (baryonic matter including compact objects accounting for less than a tenth) like WIMPS, axions, etc. There is no unambiguous detection of any of these particles. [3, 4]

In an alternate approach to dark matter, Drexler assumes that highly relativistic protons trapped in the halo of the galaxies due to the galaxies' magnetic field could possibly account for the yet unseen DM [5]. Here we look at various energetics involved in such a scenario and see why it does not seem a plausible model for DM.

The energy $(\varepsilon_P)$ of these highly relativistic protons required to be trapped in an orbit of radius ~30kpc in the galactic magnetic field of ~$10^{-6}$G is given by:

$$R = \frac{pc}{eB} \qquad \ldots (1)$$

This works out to be: $\varepsilon_P \sim 10^{16} eV$ ... (2)

To account for ~$10^{12}$ solar mass of DM in each galaxy, the number of such high energy protons required is, $n_P \sim 10^{62}$. The total number of protons in the galaxy ~$10^{67}$, therefore, one in every $10^5$ protons should be ultra-relativistic.

The total energy associated with these protons:

$$E_P = n_P \times \varepsilon_P \approx 10^{66} \, ergs \qquad \ldots (3)$$

The only source of such high energy protons is supernova explosions. Each SN generates an energy of $E_{SN} \approx 10^{51} \, ergs$

Therefore the number of SN required to produce the required number of these protons is:

$$N_{SN} = \frac{E_P}{E_{SN}} \sim 10^{15} \qquad \ldots (4)$$

This, over the average lifetime of the galaxy of ~$10^{10}$ years, implies about $10^5$ SN/year, which is much (a million times!) above the observed limit.

If it is suggested that those ultrahigh energy protons originated at early epochs of the universe, then their energies at say z~$10^9$, when the ambient background temperature was $10^{10}$ K, would have been of ~$10^{13}$ ergs (i.e. ~$10^{25}$ eV!). This energy would have red-shifted to $10^{16}$ eV at present. Buts as we see subsequently their energy losses (from scattering with the background radiation) would have scaled as $(1+z)^6$.

The high energy protons will interact with the magnetic field of the galaxy and emit synchrotron radiation of frequency given by:

$$\omega_{Syn} = \frac{\gamma^2 eB}{2\pi m_P c} \approx 10^{11} \, Hz \qquad \ldots (5)$$

Where, $\gamma = \frac{\varepsilon_P}{m_P c^2} \approx 10^7$.

The energy loss due to the synchrotron radiation is given by:

$$\dot{E}_{Syn} = \sigma_T \left(\frac{B^2}{8\pi}\right) \gamma^2 c \approx 2 \times 10^{-14} \, ergs/s \qquad \ldots (6)$$

The lifetime of the protons due to the synchrotron loss is given by:

$$\tau_{Syn} = \frac{\varepsilon_P}{\dot{E}_{Syn}} \approx 10^{18} \, s \qquad \ldots (7)$$

These high energy protons will also lose energy due to inverse Compton (IC) radiation due to interaction with the cosmic microwave background radiation (CMBR). The energy loss due to IC is given by:

$$\dot{E}_{IC} = \sigma_T (aT^4) \gamma^2 c \approx 2 \times 10^{-19} \, ergs/s \qquad \ldots (8)$$

And the corresponding lifetime is given by:

$$\tau_{IC} = \frac{\varepsilon_P}{\dot{E}_{IC}} \approx 10^{23} \, s \qquad \ldots (9)$$

At higher redshifts the energy loss will be higher. The synchrotron loss (from equation (6)) goes as $\gamma^2$ and IC loss (from equation (8)) goes as $T^4$ and both gamma factor and temperature has a dependence on the redshift, and goes as $(1+z)$. Therefore the total energy loss at higher redshifts will be higher by a factor of $(1+z)^6$ and lifetime decreases by the same factor. Therefore at a redshift of ~10, the lifetimes of these high energy protons will be ~ $10^{17} s$, which is less than the Hubble age of the universe. Therefore these protons would not have lasted till the current epoch.

To account for the missing mass in each galaxy, there should be ~$10^{62}$ relativistic protons in the halo of each galaxy. This excess charge around the galaxy will cause a tremendous Coulomb repulsion between them (eleven orders greater than their gravitational attractions).

For the gravitational attraction between Milky Way and the Andromeda galaxies to dominate, the maximum charge is constrained to be:

$$Q = \sqrt{G} M \approx 10^{51} e \qquad \ldots (10)$$

which is ~$10^{11}$ orders smaller than the number of relativistic protons required to account for the dark matter.

In order to overcome this excess charge there should be an equal number of anti-protons. But this will result in an annihilation of proton and antiproton, at a rate of ~ $n^2 \sigma v \approx 10^{32} / cm^2 / s$
Where, the number density, $n \approx 10^{-6} / cc$; $v \approx c$; $\sigma \approx 10^{-30} cm^2$ (1 micro-barn)

That gives $\sim 10^{32}$ annihilations per second producing $\sim 10^{40} ergs/s$ of high energy gamma rays. This is much higher than the observed high energy gamma ray flux.

The above are suggestive arguments against this hypothesis. Several other reasons can be given which render this model of DM untenable. So in conclusion, this alternate scenario (Drexler DM) is unfeasible.

**Reference:**

1. Perlmutter S. *et al*, Nature 391, 51, 1998
2. Krauss L. M., Ap. J., 604, 91, 2004
3. Zioutas K. *et al*, New J. Phys., 11, 105020, 2009
4. Report on PASCOS meeting, CERN Courier, p.23, Nov. 2010
5. Drexler J., *Our Universe via Drexler Dark Matter* (Universal-Publishers), Nov. 2009; Drexler J., *Comprehending and Decoding the Cosmos* (Universal-Publishers), May 2006; Drexler J., *Discovering Postmodern Cosmology* (Universal-Publishers), March 2008